# Observed and estimated prevalence of Covid-19 in Italy: Is it possible to estimate the total cases from medical swabs data?

F. Bassi – Department of Statistical Sciences, University of Padova, Italy,

G. Arbia, Department of Statistical Sciences, Catholic University of the Sacred Hearth, Milano, Italy,

P.D. Falorsi, Italian National Statistical Institute.

***Abstract:*** *During the current Covid-19 pandemic in Italy, official data are collected with medical swabs following a pure convenience criterion which, at least in an early phase, has privileged the exam of patients showing evident symptoms. However, there are evidences of a very high proportion of asymptomatic patients (e. g. Aguilar et al., 2020; Chugthai et al, 2020; Li, et al., 2020; Mizumoto et al., 2020a, 2020b and Yelin et al., 2020). In this situation, in order to estimate the real number of infected (and to estimate the lethality rate), it should be necessary to run a properly designed sample survey through which it would be possible to calculate the probability of inclusion and hence draw sound probabilistic inference. Some researchers proposed estimates of the total prevalence based on various approaches, including epidemiologic models, time series and the analysis of data collected in countries that faced the epidemic in earlier time (Brogi et al., 2020). In this paper, we propose to estimate the prevalence of Covid-19 in Italy by reweighting the available official data published by the Istituto Superiore di Sanità so as to obtain a more representative sample of the Italian population. Reweighting is a procedure commonly used to artificially modify the sample composition so as to obtain a distribution which is more similar to the population (Valliant et al., 2018). In this paper, we will use post-stratification of the official data, in order to derive the weights necessary for reweighting them using age and gender as post-stratification variables thus obtaining more reliable estimation of prevalence and lethality.*

1.  **Introduction**

In this period of pandemic emergency, most statistical analyses on Covid-19 focus primarily on finding the best forecasting model to be able to anticipate the number of epidemic cases at a national and local level (see for a review, for example, Ceylan, 2020). Comparatively less attention has been paid in the literature to a detailed descriptive analysis of the publicly available data. In Italy, there are two major sources of the data on the diffusion of Covid-19: the Istituto Superiore di Sanità - Italian National Institute of Health (INIH) and the Dipartimento di Protezione Civile - Civil Protection Department (CPD). Both sources, at the moment, do not disclose individual data on the epidemic. In particular CPD publishes daily the number of positive patients, of deaths and of individuals tested by swabs at national, regional, and provincial level. In addition, INIH releases a biweekly report with infections and deaths disaggregated by gender and age class at a national level. Positive cases are also disaggregated by age (but not by gender) in each of the 20 Italian regions. So both sources do not provide any information about the asymptomatic patient, thus limiting the usefulness of these data in view of calculating interesting epidemic parameters such as the *lethality rate*. However, a careful look at these data, especially if compared with the dimension and the demographic structure of the actual population of our country, might give important information to better understand the effects of the virus. The major aim of this paper is to calculate the prevalence and lethality rate of the Covid-19 infection in the Italian population distinguishing by age and gender. In particular, we analyse the data published by the INIH and the CPD (henceforth the "official data") with reference to the positive patients by Covid-19 and to the number of deaths at May 7 2020. The information on the demographic structure of the Italian population, refers instead, to 1 January 2019 and it is taken from the National Statistical Institute website (www.istat.it/en)

A second and more ambitious goal of this paper is to try and estimate the prevalence and the lethality of the virus in the total Italian population, considering the individuals tested with nasopharyngeal swab as a sample. Our estimation makes use of all available information on the number of positive patients and on swabs at national and regional level. As it is well known, these data suffer from the severe limitation of being observed without a proper sample design and so they can be seen as a convenience sample that cannot be used to draw probabilistic inference. To try and reduce such distortion, we propose to post-stratify the convenience sample using gender and age as post-stratification variables in order to obtain a dataset which is closer to a representative sample of the Italian population, under some reasonable assumptions.

The rest of the paper is organized as follows. Section 2 is devoted describe the prevalence and the lethality as they emerge from the officially released data. Section 3 presents an estimation of the prevalence of Covid-19 in Italy based on the post-stratification. Section 4 concludes.

## 2. A descriptive analysis of the prevalence and deaths in Italy

Table 1 reports the absolute frequencies and percentages of infected people and of deaths for Covid-19 distinguishing by gender and age (10-year classes). It also contains the implied lethality rate measured as the ratio between the number of deaths and the number of infected people. This information, published twice a week by the INIH, shows the way in which positive cases are distributed in the various demographic groups. Table 2 reports the proportion of infected people and of deaths referring to the consistency of each class in the population measured as the ratio between the number of infected (and deaths) and the total susceptible population. In principle, the ratio between the number of infected and the total susceptible population should measure the prevalence of the virus, but on the available data the measure which is obtained is not valid in that it does not show how the virus affects the entire Italian population considering its demographic structure.

**Table 1**. Positive patients, deaths and lethality rate by Covid-19 by gender and age, 7 May 2020, official data. Source: Official data by INIH.

| Age | Positive patients | | | | Deaths | | | | Lethality rate | | |
|---|---|---|---|---|---|---|---|---|---|---|---|
| | Male | Female | Total | Proportion of males | Male | Female | Total | Proportion of males | Male | Female | Total |
| 0-9 | 871 | 759 | 1630 | 53,44% | 1 | 2 | 3 | 33,33% | 0,11% | 0,26% | 0,18% |
| 10-19 | 1454 | 1449 | 2903 | 50,09% | 0 | 0 | 0 | 0,00% | 0,00% | 0,00% | 0,00% |
| 20-29 | 4961 | 6426 | 11387 | 43,57% | 6 | 3 | 9 | 66,67% | 0,12% | 0,05% | 0,08% |
| 30-39 | 7271 | 8862 | 16133 | 45,07% | 35 | 19 | 54 | 64,81% | 0,48% | 0,21% | 0,33% |
| 40-49 | 11590 | 15908 | 27498 | 42,15% | 184 | 62 | 246 | 74,80% | 1,59% | 0,39% | 0,89% |
| 50-59 | 18047 | 20297 | 38344 | 47,07% | 778 | 215 | 993 | 78,35% | 4,31% | 1,06% | 2,59% |
| 60-69 | 17678 | 11529 | 29207 | 60,53% | 2299 | 676 | 2975 | 77,28% | 13,00% | 5,86% | 10,19% |
| 70-79 | 18355 | 13231 | 31586 | 58,11% | 5566 | 2283 | 7849 | 70,91% | 30,32% | 17,25% | 24,85% |
| 80-89 | 15784 | 22206 | 37990 | 41,55% | 6593 | 4801 | 11394 | 57,86% | 41,77% | 21,62% | 29,99% |
| 90+ | 3632 | 13334 | 16966 | 21,41% | 1556 | 2873 | 4429 | 35,13% | 42,84% | 21,55% | 26,11% |
| Unknown | 24 | 31 | 55 | 43,64% | 0 | 0 | 0 | 0,00% | 0,00% | 0,00% | 0,00% |
| Total | 99667 | 114032 | 213699 | 46,64% | 17018 | 10934 | 27952 | 60,88% | 17,07% | 9,59% | 13,08% |

**Table 2**. Proportion of infected and proportion of deaths in the Italian population by gender and age, 7 May 2020: Source: Official data by INIH.

| Age | Positive patients | | | Deaths | | |
|---|---|---|---|---|---|---|
| | Male | Female | Total | Male | Female | Total |
| 0-9 | 0,03% | 0,03% | 0,03% | 0,00% | 0,00% | 0,00% |
| 10-19 | 0,05% | 0,05% | 0,05% | 0,00% | 0,00% | 0,00% |
| 20-29 | 0,15% | 0,21% | 0,18% | 0,00% | 0,00% | 0,00% |
| 30-39 | 0,20% | 0,25% | 0,23% | 0,00% | 0,00% | 0,00% |
| 40-49 | 0,25% | 0,34% | 0,30% | 0,00% | 0,00% | 0,00% |
| 50-59 | 0,39% | 0,43% | 0,41% | 0,02% | 0,00% | 0,01% |
| 60-69 | 0,50% | 0,30% | 0,40% | 0,07% | 0,02% | 0,04% |
| 70-79 | 0,67% | 0,41% | 0,53% | 0,20% | 0,07% | 0,13% |
| 80-89 | 1,13% | 1,03% | 1,07% | 0,47% | 0,22% | 0,32% |

| 90+   | 1,74% | 2,36% | 2,19% | 0,74% | 0,51% | 0,57% |
| Total | 0,34% | 0,37% | 0,35% | 0,06% | 0,04% | 0,05% |

Percentages in the last row of Table 2 indicate that this measure of prevalence is slightly higher for females than for males. This is a rather new evidence. On April 9 2020, for example, the prevalence was higher for males (0,24% vs. 0,21%) while on April 16, it was equal for the two genders (0,29%). An explanation of this result could be that in the last days observed in our reference period, even patients with light symptoms were tested and resulted positive to the infection and women showing less severe symptoms were examined in a larger percentage. In the absence of conclusive results, it is worthwhile to monitor accurately this phenomenon over time to verify if this tendency is confirmed.

**Figure 1**. Prevalence in the Italian population by gender and age, 7 May 2020. Source: Official data by INIH

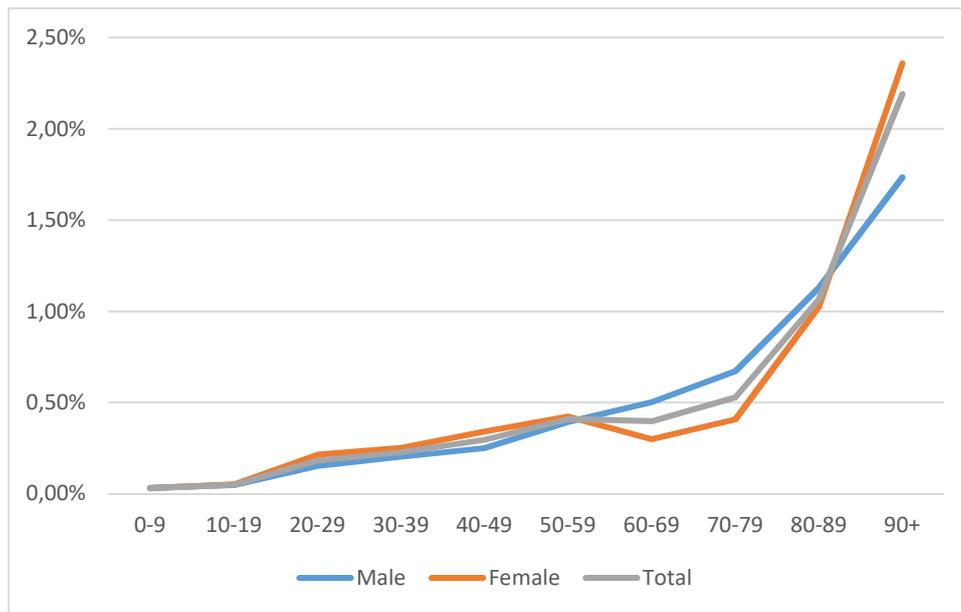

Figure 1 clearly shows that prevalence in the sample increases with age for both genders. Overall, the prevalence is higher for women than for men. More specifically, this is true for people younger than 59 years and for those older than 90, while for the other age classes we observe an opposite trend. Another interesting result emerging from the official data, is that women with an age between 60 and 80 show a lower risk of infection than women at least 10 years younger.

**Figure 2.** Proportion of deaths in the Italian population by gender and age, 7 May 2020. Source: Official data by INIH.

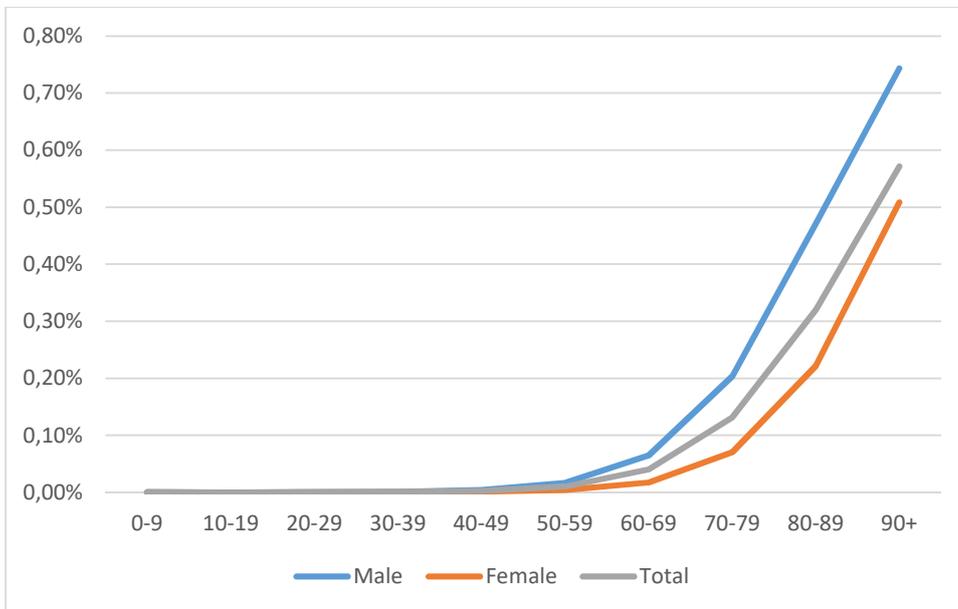

The composition indices reported in Table 3 compare the distribution of patients by gender and age with the distribution of the same variables in the Italian population. A composition lower than 1 indicates that individuals in that class are present among positive patients in a lower proportion with respect to the whole population. In contrast, a value greater than 1 provides an opposite indication. If we compare the composition indexes in the same age group between the two genders we find a confirmation of the evidences shown in Figure 1.

Finally, Figure 2 clearly shows that the proportion of deaths increases with age. In this case, the situation is much worse for males than for females in all age classes (see also composition indices in Table 3).

**Table 3**. Composition indices for patients and deaths with reference to the Italian population, 7 May 2020. Source: Official data by INIH.

|  | Patients | | Deaths | |
| --- | --- | --- | --- | --- |
| Age | Male | Female | Male | Female |
| 0-9 | 0,0981 | 0,0905 | 0,0007 | 0,0014 |
| 10-19 | 0,1438 | 0,1532 | 0,0000 | 0,0000 |
| 20-29 | 0,4553 | 0,6338 | 0,0032 | 0,0017 |
| 30-39 | 0,6023 | 0,7433 | 0,0170 | 0,0093 |
| 40-49 | 0,7438 | 1,0089 | 0,0692 | 0,0230 |
| 50-59 | 1,1621 | 1,2536 | 0,2934 | 0,0778 |
| 60-69 | 1,4845 | 0,8884 | 1,1306 | 0,3051 |
| 70-79 | 1,9844 | 1,2056 | 3,5243 | 1,2184 |
| 80-89 | 3,3336 | 3,0316 | 8,1550 | 3,8386 |
| 90+ | 5,1155 | 6,9555 | 12,8350 | 8,7770 |

## 3. Post-stratified estimation of prevalence lethality of Covid-19 in Italy

In the reference period of our analyses, various guidelines have been followed to test individuals with swab to ascertain the presence of Covid-19 infection. In the early phase of the epidemic, the World Health Organization (WHO) recommended to test only patients with at least three specific symptoms and reporting contact with the infection. In a second moment of time, since a large proportion of positive patients did not report any symptoms, these guidelines were relaxed and even people with light symptoms (or only with contact with other infected patients), were examined. Moreover, the procedures to access to swab can be different in the 20 Italian regions. These considerations, together with the evidences found in the literature of a high proportion of asymptomatic patients (e. g. Aguilar et al., 2020; Chugthai et al, 2020; Li, et al. , 2020; Mizumoto et al., 2020a, 2020b and Yelin et al., 2020), suggest that prevalence in the Italian population could be much higher than what appears from the official data.

In order to estimate the total number of infected by Covid-19, it should be necessary to run a properly designed sample survey through which it is possible to calculate the probability of inclusion and hence draw sound probabilistic inference. Such a survey has a high costs and it needs time to be realized. While waiting for the results of such a desirable survey, some researchers proposed various approaches to estimate the prevalence. They are based on epidemiologic models, time series models and the analysis of data collected in countries that faced the epidemic before like for Italy and Europe countries, China and Korea (see, e. g., Brogi et al., 2020).

In this paper, we propose to try and approximate the true prevalence of Covid-19 in Italy by exploiting the official data published by the INIH (Istituto Superiore di Sanità, 2020a and 2020b) and the CPD, but reweighted them so as to obtain something closer to a representative sample of the Italian population. The procedure of reweighting (Valliant et al., 2018) is used when, in most cases due to measurement error, a sample is not representative of the reference population. More specifically, reweighting is a procedure to artificially modify the sample composition, in the phase of data analysis, so as to obtain a distribution which is closer to the population. In its simplest form, reweighting assigns appropriate weights to each sample unit where weights can be defined on the basis of the inclusion probability (if known) or on the basis of available information on the population. In this case, we call it post-stratification (Holt and Smith, 1979; Little, 1993). In this last case, after choosing one or more stratification variables, whose distribution is known in the population, the sample units are weighted with the ratio between the theoretical proportion in the population and the observed proportion in the sample. In this paper, we will use post-stratification to analyse the official data sample, using age and gender proportion in the Italian population as post-stratification variables.

In our analysis, in order to correct the observed sample, we need to introduce some working assumptions which are required because all disaggregated information is not available. Indeed, as already said, data by gender and age are available only for positive patients, and not for those that are negative at the swab. As a consequence, in order to simulate the characteristics of all patients tested by oropharyngeal swab till 28 April 2020, we proceeded as follows.

First of all, we considered the positive patients disaggregated into age classes in each of the 20 Italian regions, an information provided twice a week by the INIH (Istituto Superiore di Sanità 2020b). Furthermore, the total number of people subjected to pharyngeal swab was derived using the information provided by the percentage of positive tests in each region supplied by the CPD. Lacking the appropriate age disaggregation we assumed that this percentage is constant in all age classes. The simulated sample of patients subjected to swabs is then distributed into the two genders by using the sex ratios observed among positive patients (see Table 1, column 5).

Secondly, we calculated the post-stratification weights referring to the distribution by age and gender of the Italian population at the last available date which was January 1st, 2019.

Finally we re-estimate the prevalence after post-stratification.

The aggregate out-coming value from this operation is prevalence estimated equal to 11.65%, a number that, reported to the total population of Italy, reveals that 7.031.321 people could have been affected by Covid-19 in the country as of 7 May 2020. Furthermore lethality rate is also re-estimated on the post-stratified sample leading to a rate of 1,76% and the median age of positive patients of 52 years.

Table 4 reports the estimated prevalence by age in the Italian population and the consequent estimated lethality rate. Even if it is known that the vast majority of swabs were obtained from symptomatic, this estimate of prevalence partly corrects for all those patients who have not been subjected to pharyngeal swab for various reasons, mainly for presenting light symptoms (pauci-symptomatic) or for being asymptomatic.

**Table 4**. Estimated prevalence and lethality in the Italian population by age after post-stratification.

| Age | Prevalence | Lethality |
|---|---|---|
| 0-9 | 7,04% | 0,02% |
| 10-19 | 7,06% | 0,00% |
| 20-29 | 9,36% | 0,01% |
| 30-39 | 11,62% | 0,04% |
| 40-49 | 15,02% | 0,12% |
| 50-59 | 16,24% | 0,33% |
| 60-69 | 13,25% | 1,36% |
| 70-79 | 11,83% | 3,64% |
| 80-89 | 7,45% | 4,51% |
| 90+ | 1,82% | 3,85% |
| Total | 11,65% | 1,76% |

By comparing our estimation with the current estimates based on unweighted (official) data, we observe that prevalence in age classes is much higher than that calculated with official data, with the only remarkable exception of the class of people with 90 years and over (see Table 2). This result shows that there might be a much higher proportion of people infected than those measured with test, possibly due to the fact that many positive patients do not even know of their condition because they have very light symptoms or no symptoms at all (Lavezzo et al., 2002). Our estimates of prevalence have the obvious effect to produce a much lower lethality rate in all age classes (except 90 and over) with respect that calculated with the uncorrected official data.

## 4. Summary of results and concluding remarks

In this report, we analyze data on Covid-19 infections in Italy with reference to the consistency and demographic structure of the Italian population. Looking at the official data, published by the INIH and by the CPD, in general for male patients the risk of infection increases with age. In contrast, the dynamics of infection is peculiar for female patients: it increases until the age of 50, then decreases until 80 and increases again for older people. Furthermore, the observed prevalence is higher for females than for males until the age of 59. After that age women show a lower prevalence and the distance from males increases with age. Finally, until the beginning of April 2020, men represented the majority among positive patients while afterwards the proportion of women was continuously increasing. The dynamic of the infection by Covid-19 in female patients deserves further study, both in terms of observation over time and with epidemiologic and statistical analyses.

Secondly, starting from official data on Covid-19, we estimate the prevalence of the infection in the Italian population assuming that only a small proportion of patients could access to pharyngeal swab test. Lacking adequate data disaggregation (for example the number of patients subjected to swab by gender and age) we had to make some working assumptions to which the results obtained are strongly depending. However, we believe that our methodology represents a reasonable approximation while waiting for more reliable data obtained with a properly designed national sample survey and that it could be further improved if more data were made available. In particular, it would be important to have the availability of disaggregated data, at least for gender and age classes, so as to avoid to impose our restrictive hypotheses of uniformity and to obtain better estimates of prevalence and lethality in each specific group of the population. This would enable us to identify the categories that are more exposed to the risk of infection and to support a system of active surveillance also in the period of recession of the pandemic.